\documentclass[conference]{IEEEtran}
\IEEEoverridecommandlockouts
\usepackage{cite}
\usepackage{amsmath,amssymb,amsfonts}
\usepackage{algorithmicx}
\usepackage{algorithm}
\usepackage{algpseudocode}
\usepackage{graphicx}
\usepackage{textcomp}
\usepackage{xcolor}
\usepackage{booktabs}
\usepackage{caption}
\usepackage{url}

\long\def\comment#1{}

\def\BibTeX{{\rm B\kern-.05em{\sc i\kern-.025em b}\kern-.08em
    T\kern-.1667em\lower.7ex\hbox{E}\kern-.125emX}}
\begin{document}

\title{Multi-Qubit Dyadic Phase Fixing for Fault-Tolerant Quantum Compilation}

\author{

\IEEEauthorblockN{Justin Kalloor}
\IEEEauthorblockA{\textit{Department of Computer Science} \\
\textit{University of California, Berkeley}\\
Berkeley, CA \\
jkalloor3@berkeley.edu}
\and
\IEEEauthorblockN{Mathias Weiden}
\IEEEauthorblockA{\textit{Department of Computer Science} \\
\textit{University of California, Berkeley}\\
Berkeley, CA \\
mtweiden@berkeley.edu}
\and
\IEEEauthorblockN{Ed Younis}
\IEEEauthorblockA{\textit{Computational Research Division} \\
\textit{Lawrence Berkeley National Laboratory}\\
Berkeley, CA \\
edyounis@lbl.gov}
\and
\IEEEauthorblockN{\phantom{John Kubiatowicz}}
\IEEEauthorblockA{
\phantom{Department } \\ 
\phantom{Department of C} \\
\phantom{sdfs}}
\and
\IEEEauthorblockN{John Kubiatowicz}
\IEEEauthorblockA{\textit{Department of Computer Science} \\
\textit{University of California, Berkeley}\\
Berkeley, CA \\
kubitron@berkeley.edu}
\and
\IEEEauthorblockN{Costin Iancu}
\IEEEauthorblockA{\textit{Computational Research Division} \\
\textit{Lawrence Berkeley National Laboratory}\\
Berkeley, CA \\
cciancu@lbl.gov}
\and
\IEEEauthorblockN{\phantom{John}}
\IEEEauthorblockA{
\phantom{Depar} \\ 
\phantom{Dep} \\
\phantom{sdfs}}
}

\maketitle

\begin{abstract}
Fault-tolerant quantum computing requires translating application-level quantum
circuits into the Clifford+$T$ gate set, where the $T$ gate is the dominant
resource cost. Phase kickback is an ancilla-based technique that can
dramatically reduce $T$-count for rotations with dyadic angles, but has
previously been limited to highly structured circuit families. We present
Dyadic Phase Fixing (DPF), a general multi-qubit synthesis tool that
extends phase kickback to general quantum circuits. DPF uses numerical
unitary synthesis to greedily extract dyadic angle rotations from any input
circuit. Combined with a decision matrix to automatically size the final phase gradient register, our end-to-end workflow achieves up to 70\% reduction in $T$-count compared to \texttt{gridsynth} and up to 60\% compared to Repeat-Until-Success synthesis on a diverse set of benchmarks. We map these compiled circuits to a surface-code architecture to evaluate space-time volume, demonstrating up to a 60\% reduction in this metric as well. However, for some circuits and mapping strategies the two metrics diverge
significantly, demonstrating that $T$-count alone is a useful but incomplete proxy for
fault-tolerant program costs.

\end{abstract}

\section{Introduction}
\label{sec:intro}
Fault-Tolerant Quantum Computer (FTQC) compilers are a crucial component in realizing the transformational potential of quantum computing, bridging the gap between domain-science-level circuit descriptions and fault-tolerant hardware. The central challenge is that circuits generated at the domain-science level are first translated into circuits with parameterized rotation
gates such as $U3(\phi,\theta,\lambda)$ or $Rz(\theta)$. These gates
do not have a native fault-tolerant implementation and must therefore
be translated into circuits composed of Clifford gates augmented with
magic states.

The $T$ gate is the most widely studied magic-state gate in the literature and is
considered the most expensive resource in any fault-tolerant algorithm
implementation~\cite{Litinski_2019, fowler2012surface}. Consequently, $T$-count/$T$-depth minimization has become the dominant optimization objective in FTQC compilation. Because the Clifford+$T$ gate set is
discrete, all general methods produce an approximate
implementation of a circuit with parameterized rotations. Given an
error threshold $\epsilon$, the state of the art in fault-tolerant
compilation is represented by the work ({\tt gridsynth}) of Ross and
Sellinger~\cite{ross_sellinger}, which provides an optimal ancilla-free decomposition of single-qubit $Rz$ rotations with gate count scaling
as $O(\log\frac{1}{\epsilon})$. More recent methods~\cite{Weiden_2025, paradis_2024_synthetiq} propose
multi-qubit Clifford+$T$ synthesis, but their applicability and
precision remain limited. In general, these methods must solve a
two-level optimization problem, simultaneously minimizing
approximation error and $T$-gate count.

Other methods~\cite{Sanders_Combinatorial, CodyJones_2012, Chen_holistic} trade ancilla qubits for reduced
$T$-count and are tailored to specific circuit structures. Phase
kickback~\cite{kitaev2002classical} is a well-studied ancilla-based
method that operates on rotations with dyadic angles (of the form
$\frac{2m\pi}{2^k}$). Phase kickback prepares a
phase gradient state on $k$ qubits and then uses controlled addition to kick the phase back onto the data qubit. This technique can greatly reduce $T$-count compared to generic methods such as \texttt{gridsynth}, but the restricted set of target angles limits its usage. 

This paper makes the following contributions:

We introduce a general method for synthesizing phase kickback
  circuits for arbitrary multi-qubit unitaries. The core idea is to
  minimize the number of arbitrary rotation angles and replace them
  with dyadic angle rotations. We then construct a multi-objective decision matrix that automatically selects the optimal phase
  gradient register size $k$, trading start-up overhead against
  $T$-gate savings while respecting user-specified hardware
  constraints. This automation ensures that phase kickback overhead is
  always effectively amortized, and our compiler is guaranteed in
  terms of $T$-count never to perform worse than standard single-qubit
  rotation synthesis.

Secondly, we evaluate phase kickback synthesis across a broad set of
  algorithms, including Hamiltonian Simulation, QFT, QAE, QAOA, and
  QPE~\cite{qiskit, rahman_2022, farhi_quantum_2014, cerezo_challenges_2022}, and demonstrate reductions of up to
  \textbf{70\%} in $T$-count compared to
  \texttt{gridsynth}, and by as much as \textbf{60\%} when compared against Repeat-Until-Success (RUS)~\cite{bocharov_2015_rus} circuits.
  Importantly, the naive application of phase kickback does not improve performance when compared to {\tt gridsynth} (performing 10- 80\% worse), emphasizing the utility of our
  multi-qubit dyadic angle synthesis. Because the kickback register size $k$ is a configurable parameter, developers can construct further optimization passes that
  co-optimize resource reduction together with ancilla overhead.

Finally, we show that $T$-count alone is not a complete proxy for
  fault-tolerant program performance. When a logical Clifford+$T$
  circuit is mapped to a physical fault-tolerant architecture such as
  the surface code via lattice surgery, the ultimate performance
  metric is the \emph{space-time volume} of the resulting QEC
  execution schedule: the product of the number of physical qubits and
  the number of logical cycles required. A reduction in $T$-count does
  not automatically translate into a reduction in space-time volume,
  because the ancilla overhead and gate structure of the synthesized
  circuit also affect layout and scheduling. One of the central
  findings of this paper is that the relationship between $T$-count
  and space-time volume depends critically on the compilation
  technique used to map the logical circuit to physical operations,
  and that for some circuits and mapping strategies the two metrics
  diverge significantly. In Section~\ref{sec:parallel_adders}, we discuss the impact of phase kickback on the resulting circuit structure, and analyze the trade-off of using additional registers to recover parallelism at the cost of more ancilla.

Overall, this work demonstrates that phase kickback synthesis is a
powerful and general Clifford+$T$ circuit optimization method
applicable to a very large class of algorithms. Furthermore, these
results confirm that $T$-count alone is not a universally reliable
proxy for fault-tolerant program performance, and should be evaluated
together with the space-time volume of the QEC-mapped circuit. In
Section~\ref{sec:disc}, we discuss these trade-offs and suggest new
compiler research directions.

\section{Background}
\label{sec:back}

Fault-tolerant compilation relies on a Quantum Error Correction (QEC)
code to improve algorithmic fidelity. The error-correcting code
encodes one or more \emph{logical} qubit states across a ``code patch''
consisting of many physical qubits. By increasing the number of
physical qubits, one can arbitrarily reduce the logical error rate,
enabling ever-larger algorithms to run on hardware. While this
addresses the fidelity challenges of the NISQ era, QEC introduces a
new set of compiler challenges.

Most notably, QEC codes do not easily admit a universal set of logical
operators. Logical operators must be applied across entire qubit code
patches, and each error-correcting code admits a particular set of
\emph{transversal} gates that can be performed efficiently. These
transversal gates form closed finite groups and are therefore not
universal~\cite{transgate}. While many methods have been proposed to
circumvent this limitation~\cite{Duclos_Cianci_2015, Gottesman_1999,
  horsman2012surface}, the most prominent solution across many
different codes is the injection of a \emph{magic state}. These states
must be carefully cultivated or distilled, making them a significant
bottleneck for fault-tolerant algorithms. Although recent
works~\cite{gidney2024cultivation,
  vaknin2025efficientmagicstatecultivation,
  sahay2025foldtransversalsurfacecodecultivation} have dramatically
reduced the cost of generating magic states, the probabilistic nature
of cultivation, injection, and correction, along with the error
ceilings of these methods, remain the main obstacle in the compilation
of fault-tolerant circuits. The most widely studied gate set for QEC
codes consists of the \emph{Clifford} gates, which are straightforward
to implement across many different codes~\cite{steane1996qec,
  shor1995scheme, fowler2012surface}, along with the magic $T$ gate
for universality.

An important consequence of compiling to the Clifford+$T$ gate set is the absence of \emph{continuous rotation gates}. While NISQ gate sets can include parameterized gates (e.g., $Rz$, $Rx$, $U3$), a fault-tolerant compiler must efficiently \emph{approximate} these continuous rotations as sequences of discrete Clifford+$T$ gates. Although this problem is believed to be hard for general unitaries, several algorithms have been proposed to efficiently approximate $Rz$ rotations to a Clifford+$T$ circuit up to a precision error $\epsilon$~\cite{ross_sellinger, bocharov_2015_rus, Campbell_Synthillation}. In general, the $T$-count and depth of these algorithms scale as $O(\log\frac{1}{\epsilon})$. The compiler must balance approximation error carefully to generate circuits that are compact yet still produce meaningful results, by bounding the approximation error over the entire circuit.

The goal of a fault-tolerant compiler is therefore twofold: (1)~translate a given quantum algorithm to the discrete gate set prescribed by the QEC code, ensuring that the total approximation error remains below a given $\epsilon$; and (2)~minimize the resources (physical qubits, runtime) required by the final physical circuit.

\subsection{Implementing Dyadic $Rz$ Gates}
\label{sec:pk_bg}

Phase kickback implementations of $R_z$ gates were introduced by Kitaev, Shen, and Vyalyi~\cite{kitaev2002classical} and have since been studied extensively in the context of fault-tolerant quantum compilation~\cite{CodyJones_2012, Sanders_Combinatorial}. The central idea is to prepare an ancilla \emph{phase gradient state}
\begin{equation}
    \frac{1}{\sqrt{2^k}}\sum_{j=0}^{2^k-1} e^{i\pi j/2^k}\,|j\rangle
\end{equation}
and then apply a controlled-addition operation that kicks the desired phase back onto the control qubit, effectively implementing an $R_z$ rotation. When the target rotation angle is \emph{dyadic}:
\[ Rz(\theta), \quad \theta \in \Big\{\frac{2m\pi}{2^k} \ | \ m, k \in \mathbb{Z} \Big\} \]
the adder circuits of~\cite{Gidney2018halvingcostof} make this approach dramatically more $T$-efficient and introduce no approximation error (beyond the error in the phase gradient state itself). This approach requires $k$ qubits to hold the phase gradient state and an additional $k-2$ scratch qubits for the optimal adder circuit, for a total ancilla overhead of $2k-2$ qubits.

\begin{figure}[t]
    \centering
    \includegraphics[width=0.95\linewidth]{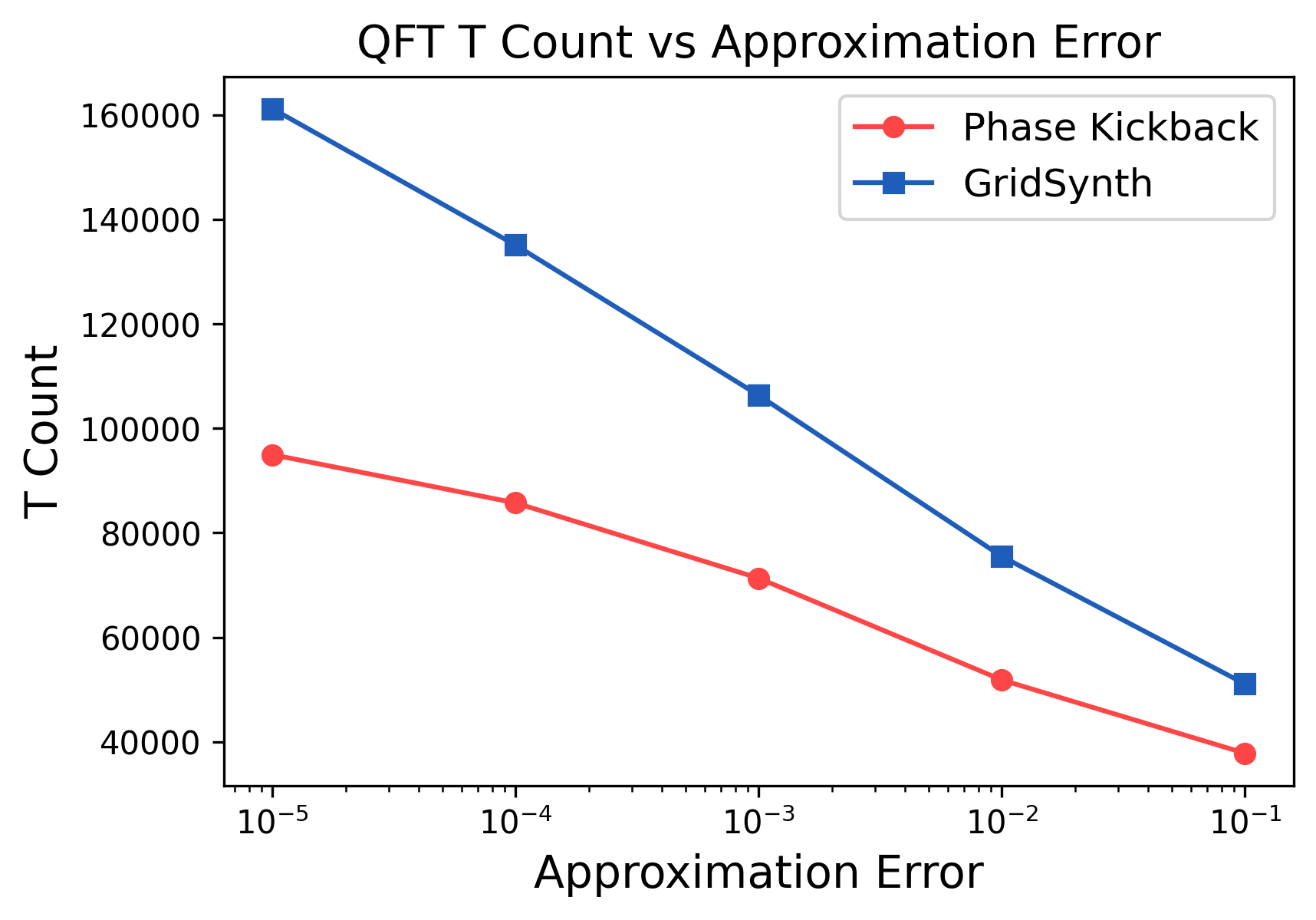}
    \vspace{-6pt}
    \caption{$T$-gate counts for the Quantum Fourier Transform circuit, comparing $Rz$ decomposition via Phase Kickback against standard rotation synthesis (\texttt{gridsynth}). Phase kickback achieves significantly lower $T$-counts for angles of the form $\frac{2\pi}{2^k}$, with the advantage growing as the approximation error threshold decreases.}
    \label{fig:qft_t_counts}
    \vspace{-16pt}
\end{figure}

As illustrated in Figure~\ref{fig:qft_t_counts} for circuits based on the quantum Fourier transform—whose $R_z$ gates are precisely of this dyadic form—phase kickback can yield substantial reductions in $T$-gate count compared to synthesizing each rotation independently.

However, this efficiency does not extend to arbitrary rotation angles. For a generic angle $\theta$, the compiler must find a value of $k$ such that $\|Rz(\frac{2m\pi}{2^k}) - Rz(\theta)\| \leq \epsilon$. A large $k$ value corresponds to a more expensive adder circuit ($T$-gates scale roughly as $4k$~\cite{Gidney2018halvingcostof}) and a larger phase gradient state. In its naive form, this restricts the technique to highly structured circuit families~\cite{Sanders_Combinatorial, Chen_holistic}.

Generalizing phase kickback to multi-qubit arbitrary circuits therefore requires methods that can rewrite the set of $Rz$ rotation angles in a circuit into a combination of dyadic and arbitrary-angle rotations. This is a two-level optimization problem: the $T$-gate count must decrease while the ancilla overhead of phase kickback remains bounded.

\subsection{Unitary Synthesis}
\label{sec:un_synth}

Unitary synthesis aims to generate a circuit $C$ whose underlying unitary $U(C)$ satisfies $d(U(C), V) \leq \epsilon$ for some target unitary $V$, distance function $d$, and error threshold $\epsilon$. Because unitary synthesis inherently relies on approximation, it is well suited to fault-tolerant compilation. Prior work~\cite{ntro} has demonstrated the power of multi-qubit approximation in the fault-tolerant setting, which we similarly leverage in our compilation workflow.

We use the BQSKit~\cite{bqskit} unitary synthesis infrastructure to synthesize our fault-tolerant circuits. BQSKit takes an ansätz circuit with \emph{continuous} rotation gates and numerically optimizes the parameters to minimize $d(U(C), V)$. If this distance falls below a configurable threshold, the circuit is accepted; otherwise, a new ansätz is generated. Because this process is computationally and memory intensive, BQSKit first partitions large circuits into smaller blocks that can be directly synthesized. Convergence is determined using the Hilbert-Schmidt distance
\begin{equation}
    d_{\text{HS}}(U, V) = \frac{\|U - V\|^2}{\sqrt{2^n}},
\end{equation}
which has the useful property that under sequential composition the total Hilbert-Schmidt distance is bounded by the sum of the per-partition distances~\cite{gilchrist_distance_2005}, allowing the approximation error across the entire algorithm to be bounded to a desired $\epsilon$.

For fault-tolerant computation, all non-Clifford gates must then be translated into a fault-tolerant gate set. The most generic approach decomposes single-qubit unitaries into $\sqrt{X}$ (Clifford) and $R_z(\theta)$ (non-Clifford) gates. For a given precision $\epsilon$, translating $R_z$ operations into Clifford+$T$ requires $O(\log \frac{1}{\epsilon})$ gates. $R_z$ rotations can be approximated \emph{optimally} in the ancilla-free case to arbitrary precision using the \texttt{gridsynth} algorithm~\cite{ross_sellinger}. Ancilla-based methods can generate more resource-efficient circuits than \texttt{gridsynth}~\cite{landahl_2013_cisc, bocharov_2015_rus, bocharov_2015_fallback}, and advances in unitary synthesis promise to discover additional such methods.

Incorporating multi-qubit ancilla-based synthesis methods into compilers that partition circuits for scalability introduces additional optimization challenges. Each block is independently synthesized with additional ancillae. When composing the final circuit, the compiler must either run ancilla reuse algorithms to bound qubit count growth, or serialize block execution. The former techniques are not well investigated, while the latter limits parallelism and increases execution time.

\subsection{Surface Codes and Lattice Surgery}

Surface codes~\cite{fowler2012surface} are one of the most promising and well studied error correction codes in the literature. A FTQC based on surface codes cannot directly execute Clifford+$T$ circuits; instead, lattice surgery~\cite{horsman2012surface} provides an efficient method for executing these programs. With lattice surgery, Clifford+$T$ circuits are translated into a series of multi-qubit Pauli Product measurements. This translation can be performed gate-by-gate, or using compilers~\cite{Litinski_2019} that commute all Clifford gates to the end of the circuit, forming higher weight Pauli Products. Executing this final series of Pauli Products is known as Pauli-Basis Computation (PBC).

\begin{figure}[tb]
    \centering
    \includegraphics[width=0.4\textwidth]{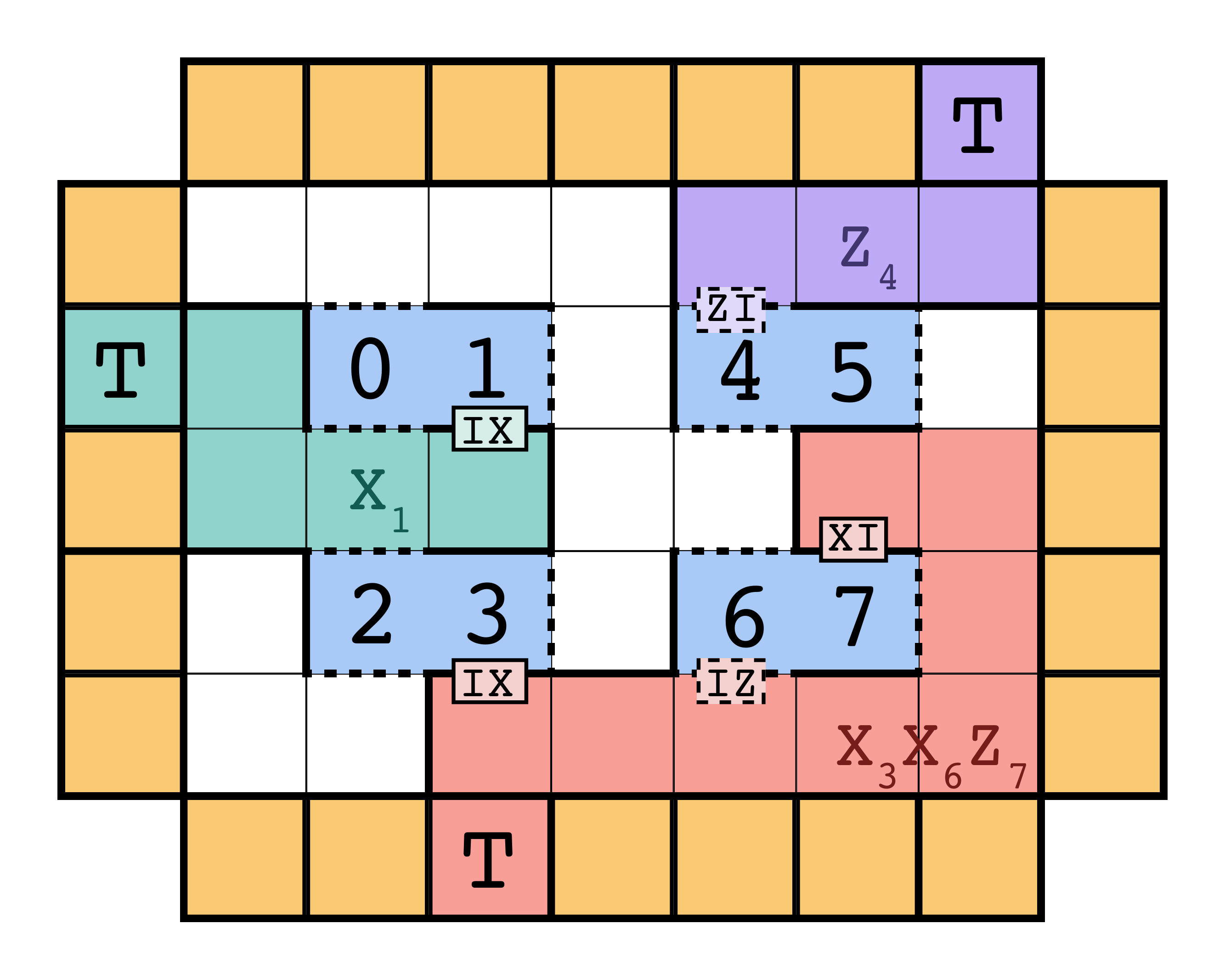}
    \caption{
    Standard bus architecture for surface code qubit patches (\cite{Litinski_2019, hofmeyr2026, silva2024lssp}). The routing qubits (blank squares) allow for logical all-to-all connectivity. Pauli Products (red, green, and purple regions) in the same cycle can be scheduled in parallel if there are non-overlapping paths through the routing qubits. The perimeter contains magic state cultivation qubits (yellow) which generate resource states as per~\cite{gidney2024cultivation}. When the resource states have a $T$ state ready, they can be used in a Pauli measurement. The final space-time volume of the mapped circuit is the number of logical cycles $\times$ all of the qubits in the architecture (data, routing, and resource).
    }
    \label{fig:topo_abstraction}
    \vspace{-20pt}
\end{figure}

Lattice surgery mappers take a PBC circuit as input and generate the physical execution schedule. To perform the required lattice surgery primitives, these mappers introduce additional routing ancillae. The space-time volume of the resulting QEC schedule is the ultimate performance metric for fault-tolerant program execution. Figure~\ref{fig:topo_abstraction} illustrates an example of mapping Pauli Products on a planar architecture.

\section{Phase Kickback Synthesis}
\label{sec:synth}

\begin{figure}
    \centering
    \includegraphics[width=0.95\linewidth]{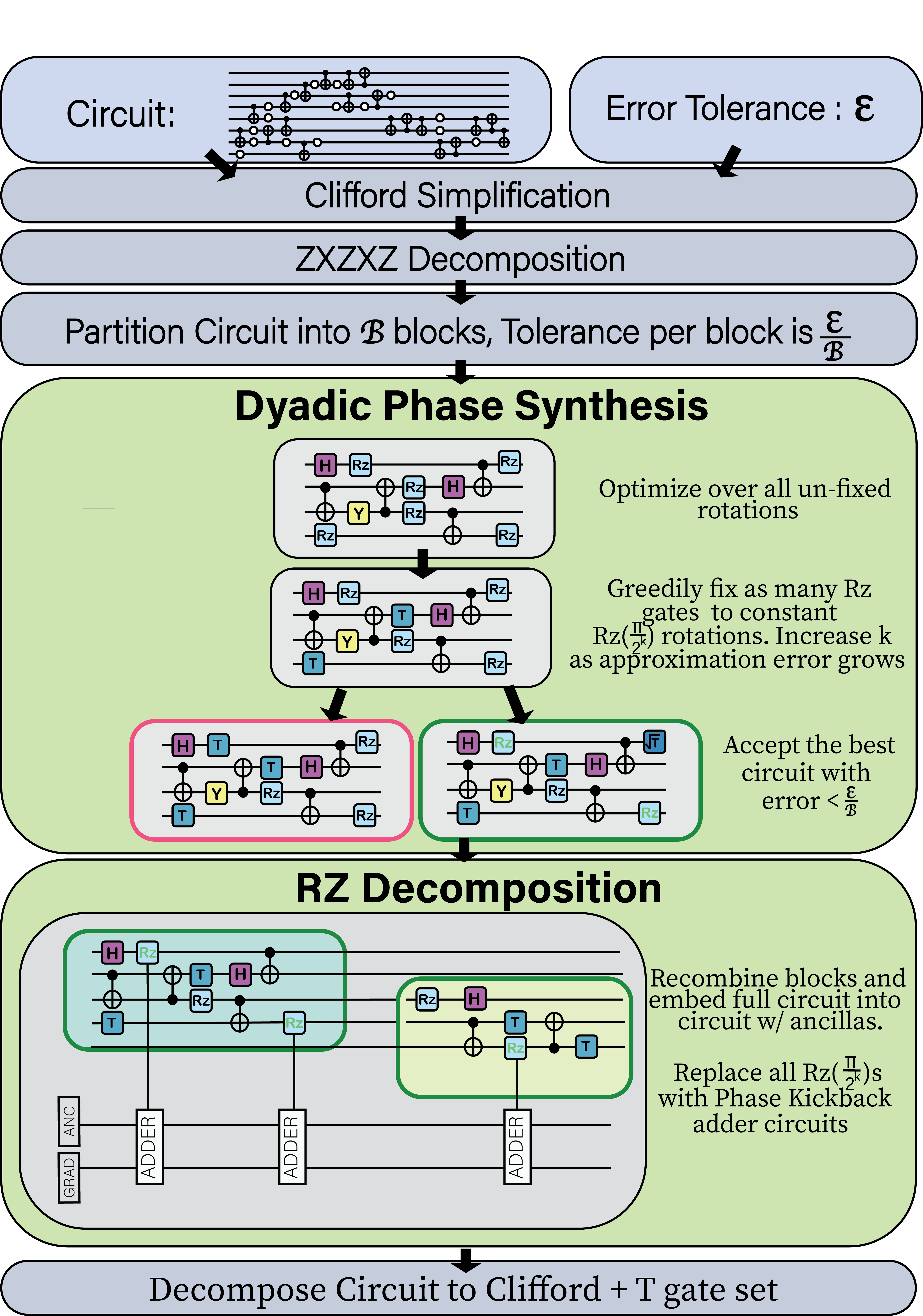}
    \caption{High-level overview of our end-to-end compilation workflow. We accept a logical circuit in any gate set along with an acceptable approximation error tolerance. The circuit is partitioned into blocks and each block is simplified with BQSKit-FT~\cite{bqskit}. Each block is then passed to our Dyadic Phase Fixing routine, which greedily extracts dyadic $Rz$ rotations from the circuit. The resulting circuit is decomposed to the Clifford+$T$ gate set and can be mapped to a given QEC code and architecture.}
    \label{fig:comp_workflow}
    \vspace{-16pt}
\end{figure}

Our end-to-end compilation workflow is shown in Figure~\ref{fig:comp_workflow}. The compiler accepts any logical quantum circuit together with an acceptable error rate for the entire algorithm, and outputs an optimized circuit in the Clifford+$T$ gate set ready for mapping to a target architecture.

Full fault-tolerant compilation necessarily requires \emph{approximation}, since continuous rotation gates must be compiled to a discrete set. Our compiler therefore accepts an input logical circuit written in any basis gate set along with an error tolerance appropriate for the given algorithm. As the error budget tightens, the resources required by the resulting circuit grow. The workflow proceeds in three stages: (1)~re-target the initial logical circuit to a gate set consisting of Clifford gates and continuous $Rz$ gates; (2)~apply our Dyadic Phase Fixing (DPF) subroutine to collapse as many $Rz$ gates as possible to multiples of $\frac{2\pi}{2^k}$; and (3)~determine whether phase kickback will yield a shorter circuit, and decompose the resulting $Rz$ gates to the Clifford+$T$ gate set.

As noted in Section~\ref{sec:un_synth}, the computational cost of numerical optimization scales exponentially with the number of qubits. To address this, the first step in our workflow is to partition the input circuit into non-overlapping sub-circuits (blocks) of width $w$ qubits. This allows our numerical optimization routines to scale to circuits with hundreds of logical qubits. Increasing $w$ can lead to higher-quality circuits at the cost of additional compile time. For all results in this paper, we use a block width of $w = 4$ qubits, which provides high-quality results with reasonable compile times.

To carefully bound the algorithmic error, we distribute the input error tolerance $\epsilon$ across all blocks. If there are $B$ blocks after partitioning, each block receives an error budget of $\frac{\epsilon}{B}$ and is processed independently. The total Hilbert-Schmidt distance is bounded by the sum of per-block errors, guaranteeing $\epsilon_T \leq \sum_{b=1}^{B} \frac{\epsilon}{B} = \epsilon$~\cite{gilchrist_distance_2005}. We note that the error budget need not be split evenly: allocating more budget to certain partitions can lead to significantly better reductions in those blocks, at the cost of tighter budgets elsewhere. Investigating adaptive error allocation strategies is left as future work.

\subsection{Dyadic Phase Fixing}

After pre-processing each block with BQSKit-FT, each partition consists of width-4 sub-circuits composed of Clifford gates and $Rz$ rotations. We now introduce our Dyadic Phase Fixing (DPF) algorithm, which runs independently on each block.

DPF has one parameter, $k_{\text{max}}$, which specifies the maximum register size allotted to phase kickback. A larger $k_{\text{max}}$ allows finer-granularity rotations, since any $Rz$ angle that is a multiple of $\frac{2\pi}{2^k}$ can be expressed exactly. Note that $k = 3$ corresponds to $Rz(\frac{\pi}{4}) = T$ gates, so $k_{\text{max}} \geq 4$ is required to observe any performance gains from phase kickback. The algorithm is described in Algorithm~\ref{alg:phase_algorithm}.

\begin{algorithm}
    \caption{Dyadic Phase Fixing($C, \frac{\epsilon}{B}$)}
    \label{alg:phase_algorithm}
    \begin{algorithmic}[1]
        \State $\mathcal{A} \gets \{C\}$, $V \gets U(C)$
        \For{$\theta \in \{\pi, \frac{\pi}{2} , \ldots, \frac{2\pi}{2^{k_\text{max}}}\}$}
            \For{$C_i \in \mathcal{A}$}
                \State $\Phi \gets \{\phi : R_z(\phi) \text{ is unfixed in } C_i\}$
                \While{$\Phi \neq \emptyset$}
                    \State $(\phi^*, m^*) \gets \operatorname{argmin}_{\phi \in \Phi,\, m \in \mathbb{Z}^+} \left|\phi - m\theta\right|$
                    \State $C_i'$ $\gets \text{Fix } R_z(\phi^*) \to R_z(m^*\theta)$
                    \State $\text{Instantiate}(C_i', V)$
                    \State $d \gets d_{\text{HS}}(U(C_i'), V)$
                    \If{$d \leq \frac{\epsilon}{B}$}
                        \State $\mathcal{A} \gets \mathcal{A} \cup \{C_i'\}$, $C_i \gets C_i'$
                    \EndIf
                    \State $\Phi \gets \Phi \setminus \{\phi^*\}$
                \EndWhile
            \EndFor
        \EndFor
        \State $C_{\text{out}} \gets \operatorname{argmin}_{C_i \in \mathcal{A}} T\text{-count}(C_i)$
        \State \Return $C_{\text{out}}$
    \end{algorithmic}
\end{algorithm}

The algorithm accepts an input circuit $C$ and a target error $\frac{\epsilon}{B}$, and outputs a circuit $C_{\text{out}}$ satisfying $d_{\text{HS}}(C, C_{\text{out}}) \leq \frac{\epsilon}{B}$. We begin with $k = 1$, corresponding to $Rz(\theta = \frac{2\pi}{2^1}=\pi) = Z$ gates. For each $Rz$ rotation in $C$, we identify the angle $\phi^*$ closest to a multiple of $\pi$ and \emph{fix} it to $m^*\pi$ to produce a new circuit $C'$. We then use numerical synthesis to optimize over all remaining unfixed $Rz$ angles to minimize $d_{\text{HS}}(U(C'), U(C))$. If this distance is within budget, we add $C'$ to the set of accepted circuits $\mathcal{A}$ and continue greedily fixing the remaining angles. After exhausting all angles at $k=1$, we proceed to $k=2$ (i.e., $Rz(\frac{m\pi}{2})$ rotations), now considering all circuits $C_i \in \mathcal{A}$ rather than only the original $C$. This process repeats for all $k \in \{1, \ldots, k_{\text{max}}\}$.

Because the algorithm fixes angles greedily in order of proximity to the nearest dyadic multiple, it does not exhaustively search all possible combinations of angle fixings. The set $\mathcal{A}$ at the end of the routine therefore contains all circuits reachable by this greedy procedure that satisfy the distance threshold, which may not include every circuit satisfying the threshold. From this set we select the circuit that minimizes $T$-gate count. The greedy structure keeps the algorithm tractable while still achieving substantial dyadic angle extraction in practice, as demonstrated by the results in Section~\ref{sec:res}.

\subsection{Optimizing Phase Kickback}
\label{sec:pk_decision}

Following the DPF synthesis pass, the circuit blocks are rejoined to form a single circuit consisting of Clifford gates and $R_z$ gates, with a subset of the $R_z$ gates fixed to angles of the form $\frac{m\pi}{2^k}$ for various $k \in [1, k_{\text{max}}]$. The compiler must then determine the optimal configuration of the phase gradient register used to implement these rotations via kickback.

Although phase kickback reduces the per-gate $T$-cost, the overhead of preparing the phase gradient state must be amortized over all $R_z$ gates that benefit from it. If too few $R_z$ gates are fixed to multiples of $\frac{2\pi}{2^k}$, this startup cost dominates and phase kickback yields no net advantage over standard synthesis. Moreover, the per-gate savings depend on the approximation error budget allocated to each $R_z$: tighter error thresholds increase the cost of conventional synthesis while leaving phase kickback largely unaffected.

\begin{figure}[t]
    \centering
    \includegraphics[width=0.98\linewidth]{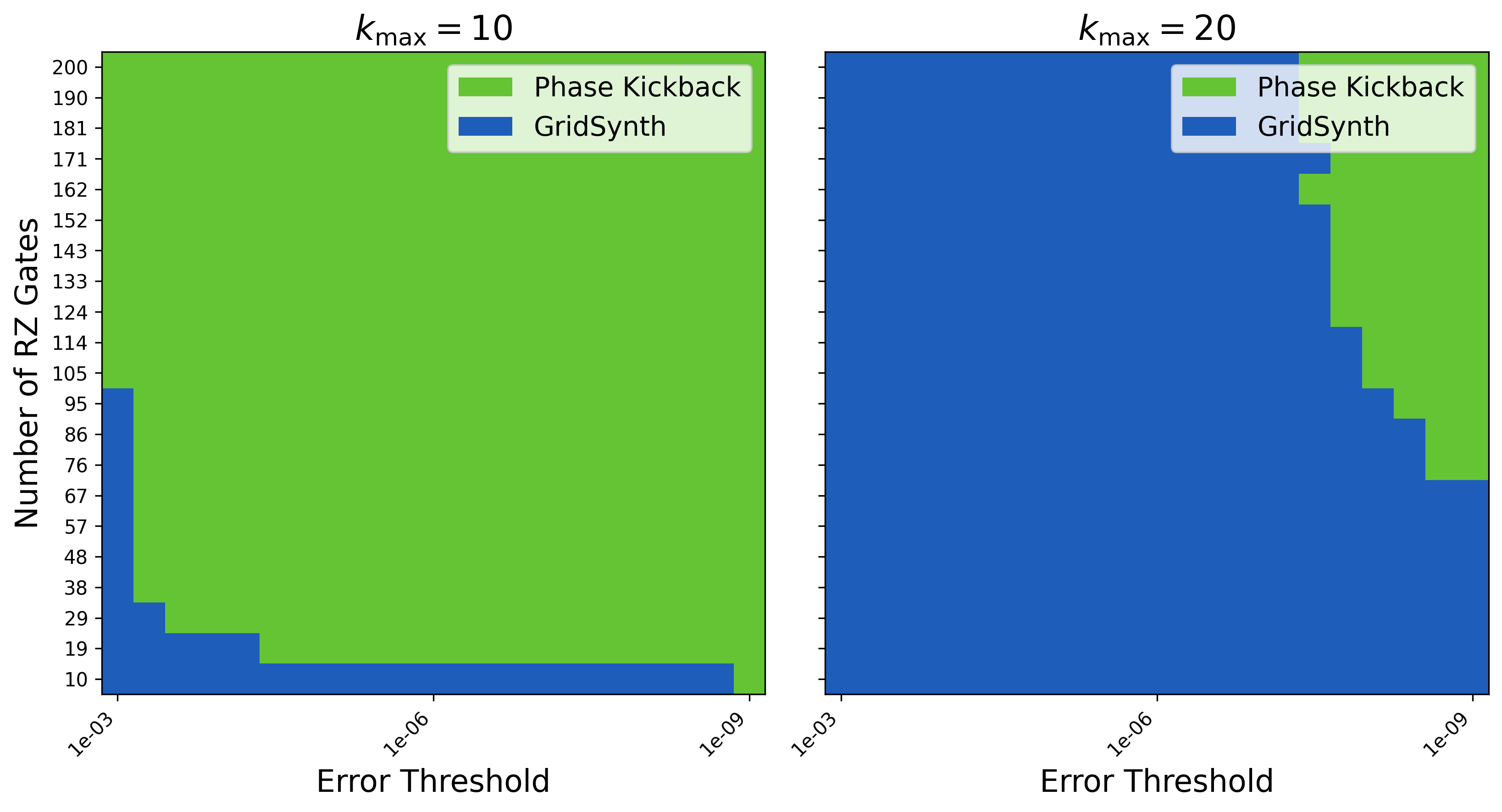}
    \vspace{-8pt}
    \caption{Decision matrix comparing $T$-gate counts between Phase Kickback and  \texttt{gridsynth} as a function of approximation error threshold and number of fixed $R_z$ gates, shown for $k=10$ (left) and $k=20$ (right). Green indicates configurations where phase kickback achieves fewer $T$ gates; Blue indicate where standard \texttt{gridsynth} is preferable.}
    \label{fig:decision_matrix}
    \vspace{-18pt}
\end{figure}

-To navigate these trade-offs automatically, our compiler constructs a three-variable decision matrix parameterized by the phase gradient register size $k$, the number of $R_z$ gates using that $k$ value, and the per-gate approximation error. The matrix compares the T-count of phase kickback against that of \texttt{gridsynth} using the analytical T-count models for both methods; the \texttt{gridsynth} model is exact, while the phase kickback model uses the adder T-count of~\cite{Gidney2018halvingcostof} together with the phase gradient state preparation cost. By sweeping over all candidate $k$ values from $4$ to $k_{\text{max}}$, the compiler selects the register size $k_{\text{reg}}$ that minimizes the overall $T$-gate count according to these models. When the decision matrix determines that no value of $k$ yields a net benefit, the compiler falls back entirely to \texttt{gridsynth}, ensuring that the compiled circuit is never worse in T-count than the baseline. We illustrate these decision frontiers for two different $k$ values in Figure~\ref{fig:decision_matrix}.

\subsection{Decomposing to Clifford+$T$}

After the decision matrix selects the optimal phase gradient register size $k_{\text{reg}}$, the final stage decomposes the full circuit into the Clifford+$T$ gate set, producing a logical circuit ready for mapping and execution on fault-tolerant hardware. The original circuit is expanded with $2k_{\text{reg}} - 2$ ancilla qubits: $k_{\text{reg}}$ qubits for the phase gradient register and $k_{\text{reg}} - 2$ scratch qubits for the efficient adder.

Following the construction of~\cite{Sanders_Combinatorial}, the $k_{\text{reg}}$-qubit phase gradient register is initialized using Hadamard gates and $R_z$ gates of the form $R_z(\frac{2\pi}{2^k})$ for $j = 1, \ldots, k_{\text{reg}}$. For each fixed $R_z$ gate of the form $R_z(\frac{m\pi}{2^k})$ in the circuit, we substitute the constant adder circuit from Figure~12 of~\cite{Gidney2018halvingcostof}, which kicks the corresponding phase back onto the target qubit via the phase gradient register. Each logical AND gate in the adder consumes one of the $k_{\text{reg}} - 2$ ancilla qubits, which are uncomputed and reused across adder calls.

After decomposing all fixed $Rz$ gates with $k \leq k_{\text{reg}}$, the circuit consists of Clifford gates, $T$ gates, and remaining $Rz$ gates (including those from the phase gradient register initialization). We distribute the error tolerance $\epsilon$ over the number of remaining $Rz$ gates and compile each to precision $\epsilon / N_{Rz}$ using \texttt{gridsynth}~\cite{ross_sellinger}. The resulting Clifford+$T$ circuit provides a complete logical description of the computation, which can then be passed to a qubit mapper to generate a full physical fault-tolerant circuit, as we demonstrate in Section~\ref{sec:results_mapping}.

This optimization pass is entirely hardware-aware. The ancilla count can be bounded by specifying $k_{\text{max}}$ in the DPF synthesis routine. The compiler is also easily extensible to other $Rz$ decomposition techniques, dyadic angle decomposition methods, and optimization objectives ($T$-gate count, $T$-depth, and CNOT count). We discuss how to target different hardware constraints further in Section~\ref{sec:disc}.

\subsection{Recombining Phase Kickback Circuits}

When large circuits are decomposed into smaller blocks, any block transformed into a phase kickback circuit contains additional ancillae, which would naively cause a qubit count explosion when the blocks are recombined into the final circuit. To bound this growth, the compiler uses a single shared phase kickback register of width $2k-2$, independent of the input circuit's data qubit count. The consequence is that the execution of all dyadic $Rz$ and phase kickback subcircuits is serialized. Alternative adders with fewer ancilla but higher $T$-counts could also be considered; in this paper we limit our exploration to a single phase gradient state and a single adder type.
 
\section{Results}
\label{sec:res}
We now demonstrate our workflow's ability to minimize $T$-gate count in the compiled logical circuit. Our workflow accepts two inputs: a logical circuit written in any basis gate set and an acceptable error threshold $\epsilon$. We bound the total unitary distance across the circuit to $\epsilon$ by carefully setting the precision of each approximation step.

Our benchmark suite comprises a diverse set of circuits, including key quantum subroutines (QAE, QPE, QFT), quantum chemistry simulations (H$_2$O, LiH, and Fermi-Hubbard), physics simulations (Ising model, Heisenberg model, neutrino oscillations, Hubbard model, and lattice gauge theory), optimization algorithms (QAOA), and quantum machine learning circuits (QML, KNN). These circuits range in width from $8$ to $420$ qubits and were generated from various domain-level generators~\cite{qiskit, rahman_2022, farhi_quantum_2014, cerezo_challenges_2022}. We first \emph{pre-optimize} each circuit using PyTKet's Full Peephole Optimization pass~\cite{tket}, producing an optimized circuit consisting of constant entangling gates (CNOT) and continuous rotations ($U3$, $Rz$). Our compiler takes these circuits together with an algorithm approximation error as input and outputs a full Clifford+$T$ circuit. We evaluate at algorithmic error thresholds of $10^{-3}$ and $10^{-5}$.

\begin{figure*}[htbp]
    \centering
    \includegraphics[width=0.65\linewidth]{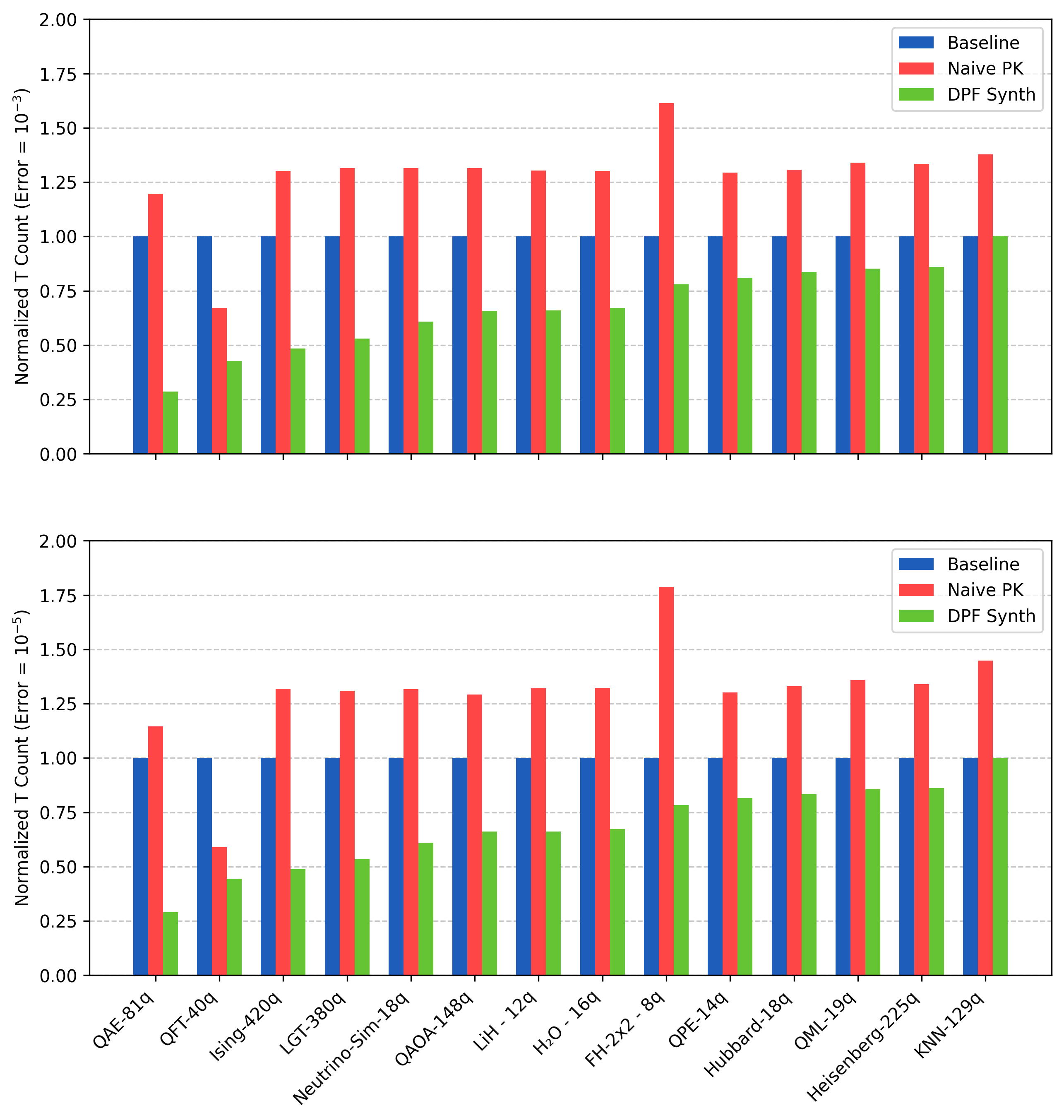}
    \vspace{-10pt}
    \caption{Normalized $T$-gate counts across all benchmarks, comparing Baseline compilation, Naive Phase Kickback (Naive PK), and Dyadic Phase Fixing Synthesis (DPF Synth) at error thresholds of $10^{-3}$ and $10^{-5}$.}
    \label{fig:t_counts}
    \vspace{-16pt}
\end{figure*}

The results are shown in Figure~\ref{fig:t_counts}. We compare three compilation strategies:
\begin{enumerate}
    \item \textbf{Baseline:} The circuit is compiled through the default BQSKit-FT pipeline, which performs Clifford simplification and collapses angles within the given $\epsilon$ to the Clifford+$T$ gate set. All continuous angles are decomposed to Clifford+$T$ using \texttt{gridsynth}.

    \item \textbf{Naive Phase Kickback (Naive PK):} All continuous $Rz$ angles are decomposed to phase kickback circuits. For each angle $\theta$, we solve for the minimum $k$ such that $\|Rz(\theta) - Rz(\frac{m\pi}{2^k})\| \leq \epsilon_{\text{gate}}$, where $\epsilon_{\text{gate}}$ is the per-gate approximation budget and $m \in \mathbb{Z}$. Each $Rz$ is then replaced by the corresponding phase kickback circuit.

    \item \textbf{Dyadic Phase Fixing Synthesis (DPF Synth):} The output of BQSKit-FT is passed through our Phase Kickback Synthesis algorithm. We automatically determine $k_{\text{reg}}$ using the decision matrix described in Section~\ref{sec:pk_decision}. For all angles of the form $Rz(\frac{m\pi}{2^k})$ with $k \leq k_{\text{reg}}$, we replace the $Rz$ gate with a phase kickback circuit. All remaining $Rz$ gates are decomposed to a Clifford+$T$ sequence using \texttt{gridsynth}.
\end{enumerate}

Across all benchmarks, our tool achieves up to a 70\% reduction in $T$-count. The largest gains are observed on Hamiltonian Simulation, QFT, QAE, QAOA, and QPE benchmarks, which tend to produce partitions with many $Rz$ gates, allowing our tool to fix several angles to multiples of $\frac{2\pi}{2^k}$ and effectively amortize the phase kickback overhead. In contrast, we observe almost no benefit for the KNN benchmark: while these circuits have many continuous parameters, our tool is unable to fix enough angles to justify the phase kickback overhead, as confirmed by the decision matrix selecting the gridsynth fallback for this benchmark. Understanding why KNN circuits prove particularly resistant to dyadic angle extraction by the greedy DPF procedure warrants further investigation.

Naive Phase Kickback is 10\% to 80\% more costly than default Clifford+$T$ synthesis across all benchmarks except QFT. This clearly demonstrates that the chosen benchmarks have \emph{not} been tailored to exploit phase kickback, and that applying the technique naively to arbitrary circuits is counterproductive. While prior work has shown the advantages of phase kickback in structured settings~\cite{Sanders_Combinatorial, CodyJones_2012, Chen_holistic}, our technique is fundamentally different: we numerically extract phase kickback-friendly angles from \emph{any input circuit} in order to minimize the final $T$-count, without assuming any special circuit structure.

The QFT circuit is a natural outlier. As noted in Section~\ref{sec:pk_bg}, the QFT consists almost entirely of dyadic $Rz$ angles by construction, making it ideally suited to phase kickback without any angle rewriting. This is visible in the results: even Naive Phase Kickback outperforms the default compilation by 40\%. Importantly, our decision matrix ensures that DPF Synth never selects phase kickback in a configuration where it increases T-count relative to gridsynth, so DPF Synth matches or improves upon the better of the two alternatives on every benchmark. While the decision matrix in this paper is constructed to optimize $T$-count against \texttt{gridsynth}, the same analysis can be performed against other $Rz$ decomposition algorithms, as we discuss in Section~\ref{sec:disc}.

\subsection{Mapping Phase Kickback}
\label{sec:results_mapping}

To obtain a more complete picture of resource costs, we map the output circuits from our end-to-end compiler to assess the impact on circuit depth and space-time volume. We focus on mapping to a logical grid of surface-code qubit patches using lattice surgery~\cite{horsman2012surface}. The grid consists of logical data qubits and logical ancilla routing qubits that provide all-to-all connectivity, with magic state cultivation patches placed around the perimeter to generate resource states (Figure \ref{fig:topo_abstraction}). We evaluate two techniques to translate our circuits to a sequence of Pauli Products that can be scheduled on chip:

\begin{enumerate}
    \item \textbf{Lightweight Pauli Basis Computation (LPBC):} Single-qubit Clifford gates are conjugated rightward using the rules outlined in Figure~4 of~\cite{Litinski_2019} until a $T$ gate is required. This results in $T$ state injection in the $X$, $Y$, or $Z$ basis rather than just $Z$-basis injection interspersed with Clifford operations. Because the single-qubit Clifford group has only 24 distinct members (up to global phase), a lookup table replaces the conjugated Clifford operations with a simpler sequence. The resulting conjugated sequences take the form

        \[
            C \times R_{\{x, y, z\}}\!\left(\frac{\pi}{4}\right) \times \dots \times R_{\{x, y, z\}}\!\left(\frac{\pi}{4}\right)
        \]
        where $C$ is an element of the single-qubit Clifford group. %\textcolor{red}{Mathias add a short sentence about how CNOTs are handled.}

    \item \textbf{Heavyweight Pauli-Basis Computation (HPBC):} HPBC follows a similar decomposition pipeline, but Clifford conjugation is applied to all single- and multi-qubit Clifford gates. In this process, we commute all Clifford gates to the end of the circuit, where they are absorbed into the measurement. This results in large Pauli Product Measurements that span many qubits, unlike LPBC where each Pauli Product involves at most two qubits. In general, HPBC eliminates the need to directly perform Clifford operations, vastly reducing the number of on-chip operations. However, high-weight Pauli products are harder to perform in parallel, which can become a bottleneck in circuit depth.
\end{enumerate}

Both pipelines map the resultant Pauli Products to a planar surface code architecture where all logical qubit measurements are performed with lattice surgery. The mapping algorithm of~\cite{hofmeyr2026} was used, as it empirically outperforms existing algorithms~\cite{Litinski_2019, silva2024lssp}. The resulting space-time volume results are presented in Table~\ref{tab:spacetime_results}.

\begin{table}[h]
    \centering
    \captionsetup{position=bottom}
    \setlength{\tabcolsep}{4pt}
    \begin{tabular}{@{}lrrrr@{}}
        \toprule
        & \multicolumn{2}{c}{\textbf{LPBC}} & \multicolumn{2}{c}{\textbf{HPBC}} \\
        \cmidrule{2-3} \cmidrule{4-5}
        \textbf{Benchmark} & \textbf{Baseline} & \textbf{DPF Synth} & \textbf{Baseline} & \textbf{DPF Synth} \\
        \midrule
        QAE-81q          & \textbf{59.6} & 92.6 ($+$55\%) & 163.4 & \textbf{64.7} (\textbf{$-$60\%}) \\
        QFT-40q          & \textbf{13.6} &  25.2 ($+$86\%) &  25.6 & \textbf{15.9} (\textbf{$-$38\%}) \\
        Ising-420q       &  32.8 & \textbf{14.2} (\textbf{$-$57\%}) &  32.8 & \textbf{14.3} (\textbf{$-$57\%}) \\
        LGT-380q         & 145.0 & \textbf{83.2} (\textbf{$-$43\%}) & 217.2 & \textbf{116.4} (\textbf{$-$46\%}) \\
        Neutrino-18q     &  13.5 & \textbf{8.7} (\textbf{$-$36\%}) &  10.7 & \textbf{5.7} (\textbf{$-$47\%}) \\
        QAOA-148q        &  25.3 & \textbf{16.6} (\textbf{$-$34\%}) & \textbf{31.1} &  42.4 ($+$36\%) \\
        LiH-12q          &   7.9 & \textbf{7.2} (\textbf{$-$9\%}) &   6.7 & \textbf{5.3} (\textbf{$-$21\%}) \\
        H$_2$O-16q       &  16.6 & \textbf{12.4} (\textbf{$-$25\%}) &  13.4 & \textbf{9.2} (\textbf{$-$32\%}) \\
        FH-2x2-8q        &   0.2 & \textbf{0.2} (\textbf{$-$17\%}) &   0.2 & \textbf{0.1} (\textbf{$-$20\%}) \\
        QPE-14q          & \textbf{21.7} &  34.0 ($+$57\%) & \textbf{20.8} &  37.6 ($+$81\%) \\
        Hubbard-18q      &   6.2 & \textbf{5.5} (\textbf{$-$11\%}) &   5.0 & \textbf{4.2} (\textbf{$-$15\%}) \\
        QML-19q          &   1.2 & \textbf{1.1} (\textbf{$-$14\%}) &   1.3 & \textbf{1.2} (\textbf{$-$9\%}) \\
        Heis.-225q       &  25.6 & \textbf{22.4} (\textbf{$-$12\%}) & \textbf{44.7} &  64.9 ($+$45\%) \\
        KNN-129q         &  1.8 &   1.8 ($+$0\%) &  1.4 &   1.4 ($-$0\%) \\
        \bottomrule
    \end{tabular}
    \caption{Space-time volumes (millions of logical qubit cycles) across all circuit benchmarks for Lightweight Pauli-Basis Computation (LPBC) and Heavyweight Pauli Basis Computation (HPBC). For each method, we compare the circuits outputted from Baseline and DPF Synth compilation. Bold indicates the lower space-time volume within each compilation strategy. Percentages in parentheses indicate the change from Baseline to DPF Synth; negative values indicate a reduction.}
    \label{tab:spacetime_results}
    \vspace{-12pt}
\end{table}

For many benchmarks, our $T$-gate reductions translate effectively into reductions in space-time volume, despite the potential increase in qubit count. Two factors explain this. First, the architecture already requires many qubits: $D$ data qubits require around $3D$ bus qubits for all-to-all connectivity, plus additional perimeter qubits for resource state generation. The $2k_{\text{reg}} - 2$ additional qubits introduced by phase kickback are often negligible relative to the existing layout. Second, in some cases the decision matrix determines that phase kickback is not warranted for a given circuit, so no additional qubits are introduced.

For QPE-14q, however, the qubit overhead clearly outweighs the $T$-gate savings under both LPBC and HPBC, and the baseline decomposition yields a smaller space-time volume in both cases. While our decision matrix is designed to minimize T-count, space-time volume additionally depends on the adder CNOT structure and qubit layout. Whether larger QPE circuits would benefit from phase kickback is an open question that requires evaluating the scaling of both T-count savings and ancilla overhead as a function of circuit width.

An important pattern in the results is the split between LPBC and HPBC for certain benchmarks: DPF Synth performs worse than the Baseline under one  but outperforms the Baseline under the other. Understanding this pattern requires examining how each technique interacts with the gate structure introduced by phase kickback.

Under LPBC, each two-qubit gate in the logical circuit is mapped to an individual lattice surgery operation. Phase kickback replaces $Rz$ gates with adder circuits containing many CNOT gates on the shared phase gradient register. These CNOTs are inherently sequential—each adder invocation must wait for the register to be free—which limits parallelism and increases depth. The degree to which this depth increase outweighs the T-count reduction depends on how many of the baseline T gates could themselves be executed in parallel. For circuits like QAE and QFT, the circuit consists of many parallel $Rz$ gates that translate into parallel Clifford+$T$ sequences. In this regime, replacing parallel T-gate sequences with sequential adder circuits massively increases the depth. The result is a larger space-time volume under LPBC for these two circuits.

Under HPBC, the situation reverses. HPBC commutes all Clifford gates—including the adder CNOTs—to the end of the circuit, merging them into high-weight Pauli product measurements. The cost of the adder CNOTs is therefore absorbed into a smaller number of broader operations rather than appearing as individual sequential steps. For QAE and QFT, this absorption eliminates the depth penalty of the adder circuits, and the T-count reduction translates directly into fewer Pauli product measurements, yielding the 38--60\% space-time volume reductions observed.

For QAOA and Heisenberg, the relationship is reversed: DPF Synth outperforms the Baseline under LPBC but underperforms under HPBC. QAOA and Heisenberg circuits have highly structured, spatially local gate patterns. We can analyze the the final mapped output by considering the \emph{scheduling efficiency}: the number of Pauli Products executed per logical cycle. The LPBC efficiency goes from 3.83 to 4.51 for QAOA and from 6.48 to 6.86 for Heisenberg for the Baseline circuit and DPF Synth circuit respectively. We can see here that trading some of these local \texttt{gridsynth} sequences for serialized adder circuits does not hurt the parallelism, but actually improves the efficiency! The keys here are the structural hazards present in the planar architecture. With DPF Synth, the number of T states required per cycle is reduced, which means the scheduler does not need to spend as much time waiting for these resource states to be cultivated. Additionally, for each logical cycle, the scheduler must wait for routing paths to open to place Pauli measurements that could otherwise be scheduled together. So, although the adder does reduce the parallelism in the circuit, the bottleneck for these circuits lies in the placement/number of ancilla in the hardware.

Under HPBC, however, the high-weight Pauli products generated by commuting the adder CNOTs compete for the same routing paths on the shared phase gradient register qubits, reducing the effective parallelism far below that of the baseline. This is captured by the scheduling efficiency dropping from 2.85 to 1.40 for QAOA and from 2.51 to 1.52 for Heisenberg under HPBC. The net effect is that the T-count reduction is more than offset by the loss of parallelism, increasing space-time volume under HPBC.

In summary, the LPBC/HPBC split is explained by two interacting factors: whether the baseline circuit has high or low inherent parallelism among its Clifford gates, and whether the adder CNOT overhead manifests as sequential individual operations (LPBC) or as high-weight parallel Pauli products (HPBC). The results show that neither mapping strategy uniformly dominates, and that the optimal choice depends on the structure of the specific circuit being compiled.

\section{Discussion}
\label{sec:disc}

\subsection{Collapsing the Dyadic Distribution}

\begin{figure}[htbp]
    \vspace{-8pt}
    \centering
    \includegraphics[width=0.95\linewidth]{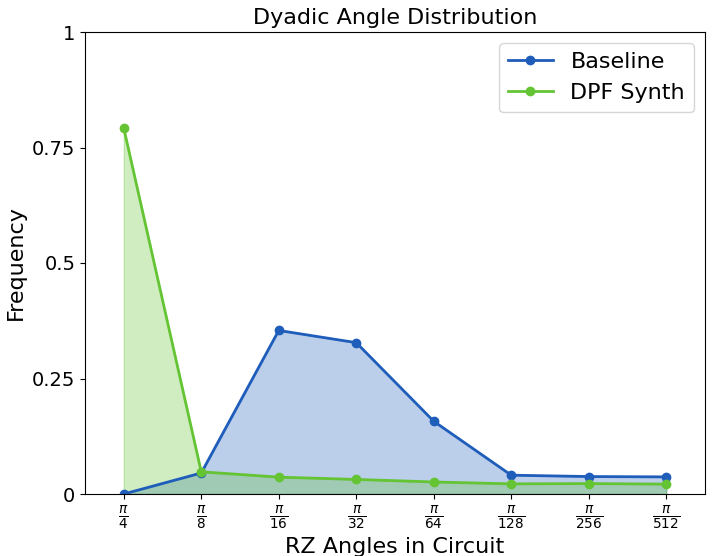}
    \caption{Comparison of dyadic $Rz$ angles across benchmark circuits that are multiples of $\frac{2\pi}{2^k}$, $k \leq 10$. The \emph{blue} distribution shows the angles found in the original circuit; the \emph{green} distribution shows the angles after Phase Kickback Synthesis. Gates that were already $T$ gates in the input circuit are excluded from both distributions, which is why the blue distribution contains no $Rz(\frac{\pi}{4})$ gates.}
    \label{fig:angle_dist}
    \vspace{-18pt}
\end{figure}

Our compiler uses decision matrices to automatically determine when to incur the overhead of phase kickback circuits. Importantly, even in cases where the decision matrix determines that full phase kickback is not warranted, our sequential greedy synthesis still achieves significant gains over the default compilation by collapsing the distribution of low-$k$ ($k \leq 10$) dyadic $Rz$ angles. This distribution shift is illustrated in Figure~\ref{fig:angle_dist}. The blue distribution shows the dyadic angles found in the original circuit (within the budgeted input error tolerance $\epsilon = 10^{-3}$), while the green distribution shows the angles after Phase Kickback Synthesis. Our method captures nearly the entire original distribution—and many additional non-dyadic $Rz$ angles—using only $T$ gates, enabling significant savings even when phase kickback circuits are not employed.

\subsection{Parallelizing Phase Kickback}

\label{sec:parallel_adders}

One of the interesting results in Table~\ref{tab:spacetime_results} is
the QAE-81q circuit under LPBC compilation. Despite achieving the
largest $T$-count reduction (70\%), it corresponds to a 55\%
\emph{increase} in space-time volume. As explained in
Section~\ref{sec:results_mapping}, this occurs because phase kickback
trades many T gates from gridsynth sequences for adder circuits that
contain fewer T gates but many CNOTs. Because the baseline QAE
circuit under LPBC has limited Clifford parallelism, the adder CNOTs
add depth without being absorbed into parallel operations.

The root cause is structural: our workflow allocates a \emph{single} phase gradient register (and associated ancilla) and replaces each dyadic $Rz$ with an adder circuit on that register. This serializes all $Rz(\frac{2\pi}{2^k})$ gates, since each adder must wait for the phase gradient register to be free. The natural remedy is to allocate multiple phase gradient registers, enabling these rotations to execute in parallel.

\begin{figure}[htbp]
    \vspace{-12pt}
    \centering
    \includegraphics[width=0.95\linewidth]{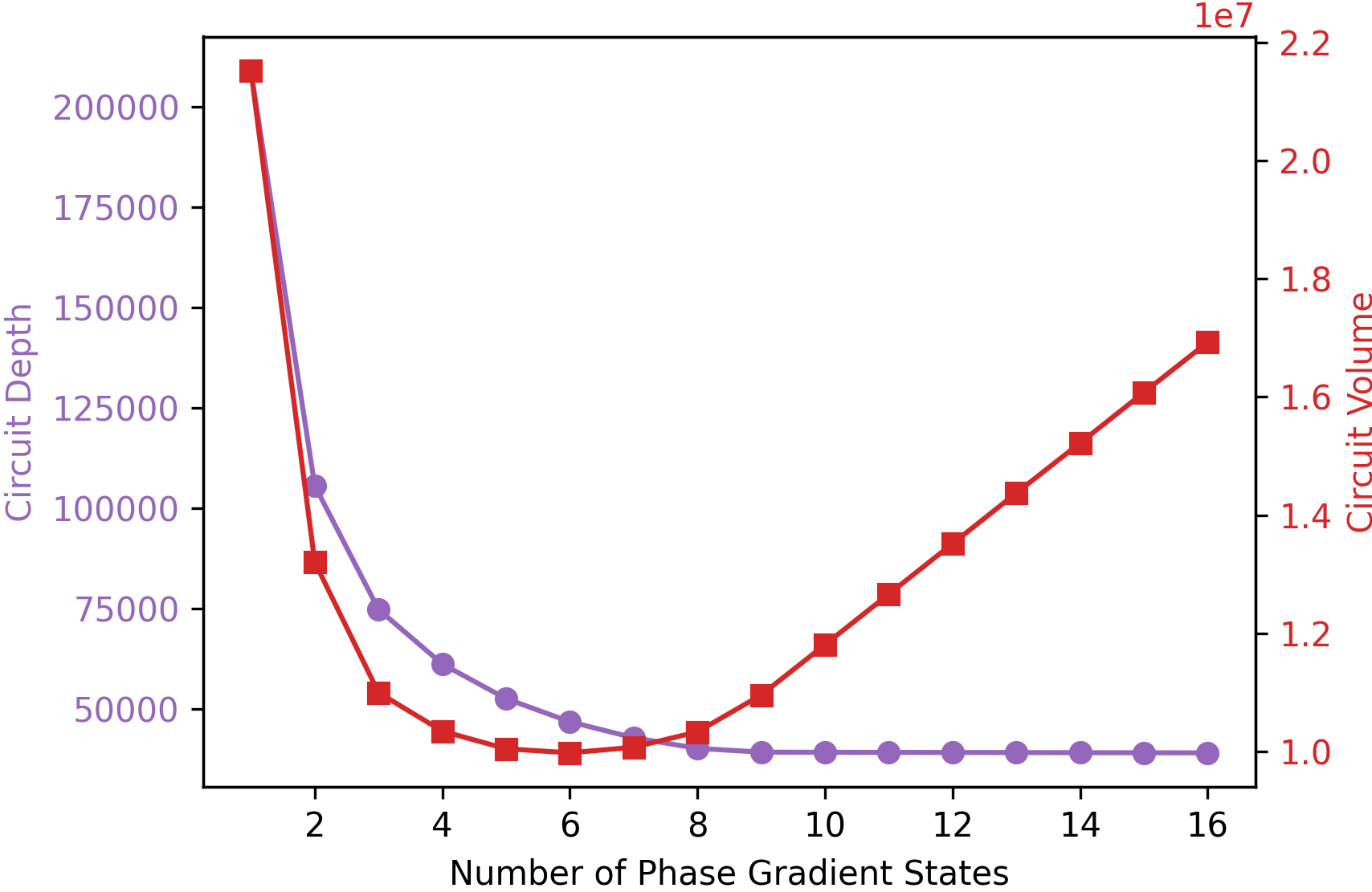}
    \caption{Impact of the number of phase gradient states on the logical circuit depth and volume of the QAE-81q circuit.}
    \label{fig:parallel_rzs}
    \vspace{-16pt}
\end{figure}

Figure~\ref{fig:parallel_rzs} plots the effect of adding additional phase gradient states on the circuit depth and volume of the QAE circuit. The adder circuits are clearly the dominant bottleneck: 75\% of the circuit depth can be eliminated by adding more phase gradient states. After six phase gradient states, however, the additional qubit cost begins to increase the space-time volume even as the depth continues to fall, indicating that the optimal operating point for this benchmark and error rate is around six registers. The appropriate number of registers will vary with circuit structure and qubit budget, and determining it automatically is a natural target for future work.

\subsection{Hardware Portability}
\label{sec:disc_rus}

\begin{figure}[h]
    \centering
    \includegraphics[width=0.5\linewidth]{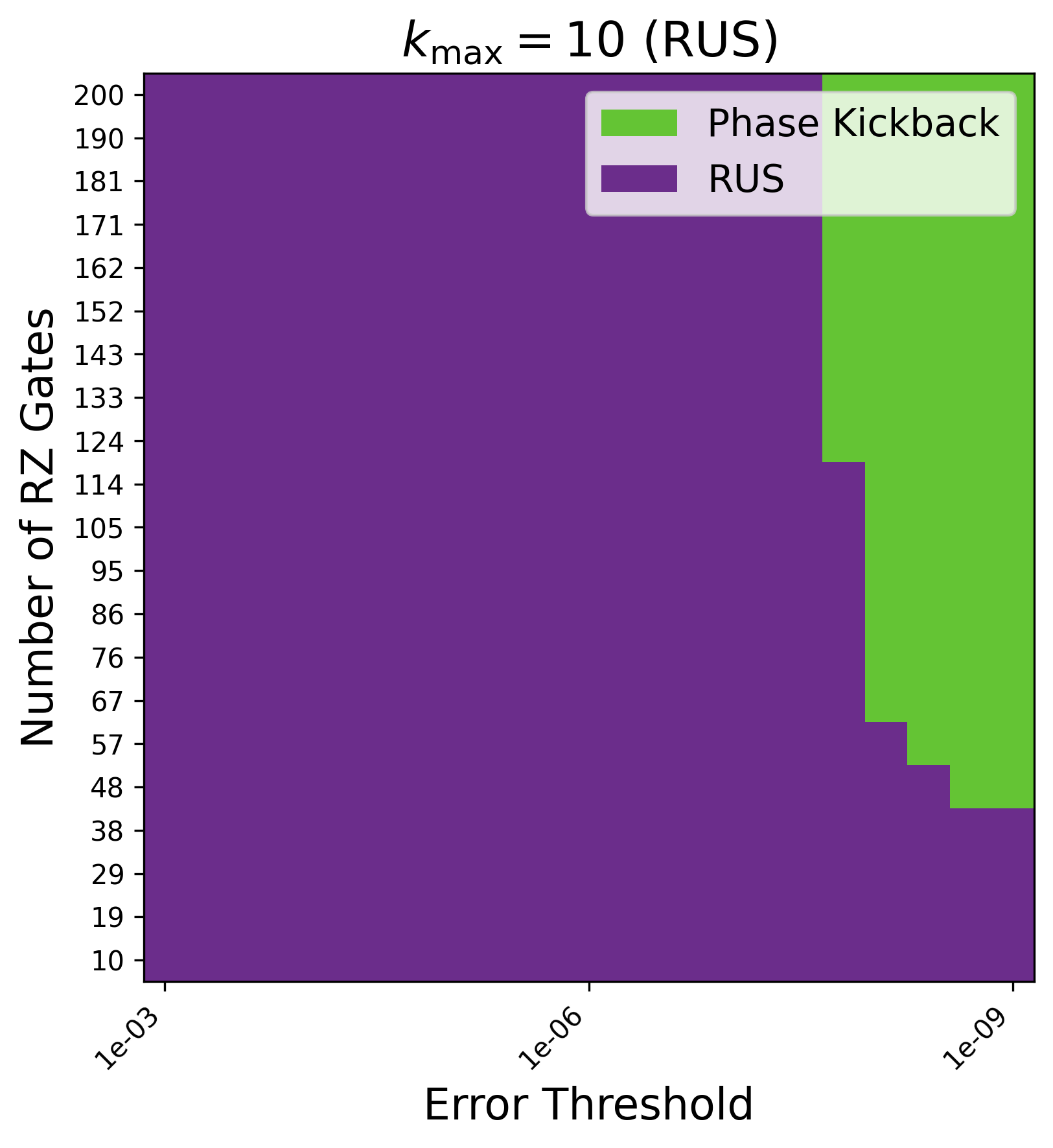}
    \caption{Decision matrix comparing Phase Kickback and Repeat-Until-Success (RUS)~\cite{bocharov_2015_rus} as a function of approximation error threshold and number of fixed $R_z$ gates. Green indicates configurations where Phase Kickback achieves fewer $T$ gates; Purple indicate where RUS is preferable. RUS requires a considerably larger number of fixed $Rz$ gates and higher precision to merit the use of Phase Kickback.}
    \label{fig:rus_decision}
    \vspace{-10pt}
\end{figure}

Our compiler is designed to be hardware-agnostic, incorporating different hardware constraints and optimization criteria with minimal changes. One important extension is support for Repeat-Until-Success (RUS) circuits, a popular technique for implementing $R_z$ gates using additional ancilla qubits that achieves lower expected $T$-gate counts than deterministic methods such as \texttt{gridsynth}. Because RUS circuits are probabilistic, their T-count is a random variable; the decision matrix for RUS is therefore built using expected T-counts from the model of~\cite{bocharov_2015_rus}, parameterized by $k_{\text{max}}$ and the per-gate approximation error. This is an approximation, and circuits with high variance in T-count may see outcomes that deviate from the decision matrix prediction. With this substitution, the rest of our synthesis pipeline requires no modification, as shown in Figure~\ref{fig:rus_decision}.

\begin{figure}[h]
    \centering
    \includegraphics[width=0.95\linewidth]{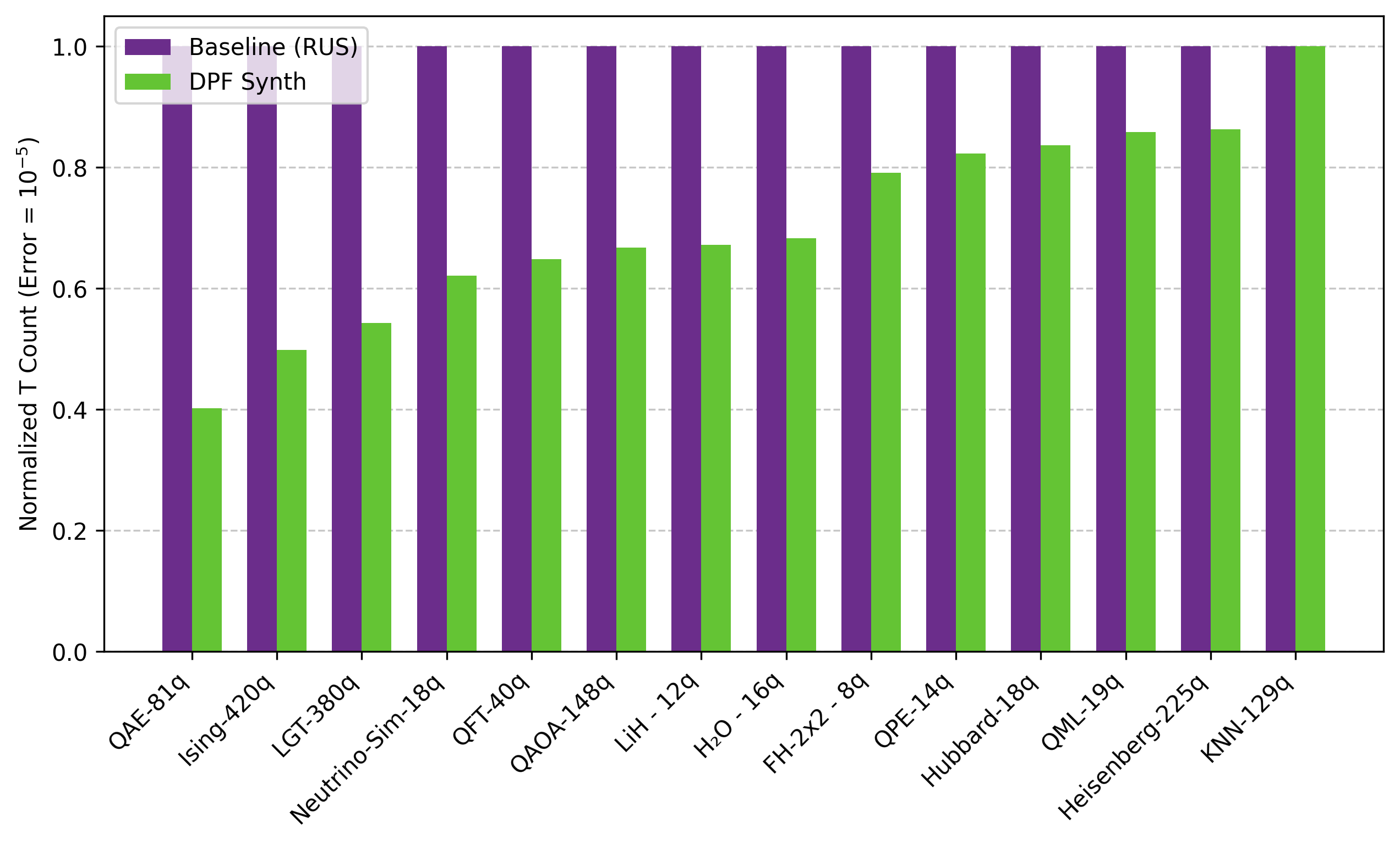}
    \vspace{-8pt}
    \caption{$T$-gate counts comparing our workflow (DPF Synth) and RUS-based rotation synthesis across benchmark circuits. DPF Synth still achieves significant reductions in $T$-gate count over the RUS baseline, demonstrating that the technique generalizes beyond \texttt{gridsynth}-based synthesis.}
    \label{fig:rus_data}
    \vspace{-14pt}
\end{figure}

Because the rest of our synthesis pipeline is hardware-agnostic, phase kickback still achieves $T$-gate count reductions over the RUS baseline when RUS is used in place of \texttt{gridsynth}, as shown in Figure~\ref{fig:rus_data}. As the RUS decision matrix shows, the crossover point at which phase kickback becomes beneficial requires more fixed $Rz$ gates and higher precision than in the gridsynth case, meaning the gains are more selective but still real for circuits with sufficient dyadic angle density.

Beyond rotation synthesis, our compiler can also operate under strict ancilla constraints. By specifying a maximum ancilla budget $k_{\text{max}}$, the compiler automatically caps the phase gradient register size to whatever the hardware can accommodate, degrading gracefully rather than failing outright when qubit resources are scarce. Looking further ahead, there is significant opportunity in automatically partitioning a given ancilla budget across competing uses: routing qubits, magic state cultivation qubits, RUS ancilla, and phase kickback registers all compete for the same physical resources, and co-optimizing this allocation is a natural next step toward a fully hardware-aware fault-tolerant compiler.

\subsection{T-Count Versus Space-Time Volume}

A consistent theme across our results is the tension between T-count and space-time volume as optimization objectives. Our compiler optimizes T-count, which is the dominant metric in the fault-tolerant compilation literature, and achieves reductions of up to 70\% on that metric. However, the mapping results in Table~\ref{tab:spacetime_results} show that these reductions translate into space-time volume improvements for most but not all benchmarks and mapping strategies, and in some cases produce significant regressions.

This divergence arises because T-count captures only one dimension of circuit cost. The adder circuits introduced by phase kickback also add CNOT gates, increase qubit count, and constrain scheduling in ways that T-count does not reflect. Whether the T-count reduction or the CNOT overhead dominates in the final space-time volume depends on the mapping strategy and the parallelism structure of the input circuit, as analyzed in Section~\ref{sec:results_mapping}.

These findings suggest that future fault-tolerant compilers should optimize space-time volume directly, rather than using T-count as a proxy. Our decision matrix framework is a natural starting point for this extension: rather than comparing T-counts, the matrix could compare predicted space-time volumes under a given mapping strategy, incorporating models for adder CNOT cost, qubit layout, and scheduling efficiency. Developing such models is a non-trivial but tractable research direction, and one that our results motivate clearly.

\section{Conclusion}
We have presented a general method for phase kickback synthesis that extends an established but narrowly applicable fault-tolerant compilation technique to arbitrary multi-qubit circuits. The key contribution is the Dyadic Phase Fixing algorithm, which uses numerical unitary synthesis to greedily extract dyadic-angle rotations from any input circuit within a specified error budget, combined with a decision matrix that automatically determines whether and how to apply phase kickback based on the circuit's rotation structure and the available ancilla budget. Together, these components ensure that the compiler never increases T-count relative to standard gridsynth-based synthesis, while achieving reductions of up to 70\% on a broad and structurally diverse benchmark suite.

The mapping results reveal a more nuanced picture. For most benchmarks and both mapping strategies evaluated, T-count reductions translate into meaningful space-time volume improvements, with gains as large as 60\% observed for QAE under HPBC compilation. However, for a subset of benchmarks the additional CNOT overhead of the phase kickback adder circuits limits or reverses this gain. The direction of the effect depends on two interacting factors: the inherent parallelism structure of the input circuit, and whether the chosen mapping strategy serializes adder operations as individual lattice surgery steps or absorbs them into high-weight Pauli product measurements. These findings demonstrate concretely that T-count is an imperfect proxy for fault-tolerant program performance, and that space-time volume under a realistic mapping strategy is the more faithful metric.

The analysis of parallel phase gradient registers shows that the space-time volume penalty of serialized adder circuits is not fundamental. For QAE-81q, allocating six phase gradient registers eliminates 75\% of the circuit depth introduced by phase kickback, recovering the space-time volume benefit that the T-count reduction predicts. This points to a general design principle: the number of phase gradient registers is a configurable parameter that trades qubit count against circuit depth, and selecting it optimally requires co-optimizing both dimensions simultaneously.

Altogether, these results identify a clear direction for future work. The decision matrix framework introduced here optimizes T-count, but its structure is equally amenable to optimizing predicted space-time volume directly, incorporating models for adder CNOT cost, qubit layout overhead, and scheduling efficiency under a target mapping strategy. Developing such models would move fault-tolerant compilation closer to the goal of directly minimizing the physical execution cost of quantum algorithms rather than a proxy for it. Phase kickback synthesis, generalized and hardware-aware, is a strong candidate for a central role in that compiler stack.

\section{Acknowledgements}
This work was supported by the U.S. Department of
Energy (DOE) under Contract No. DE-AC02-05CH11231,
through the Oﬃce of Advanced Scientific Computing
Research Quantum  Program grants FP00018850 and FP00018743.
This research used resources of the National Energy Research Scientific Computing Center (NERSC), a Department of Energy User Facility (project m306-2024).

\bibliographystyle{plain}
\bibliography{refs}

@article{shor1995scheme,
  title   = {Scheme for reducing decoherence in quantum computer memory},
  author  = {Shor, Peter W.},
  journal = {Physical Review A},
  volume  = {52},
  number  = {4},
  pages   = {R2493},
  year    = {1995},
  month   = {Oct},
  publisher = {American Physical Society}
}

@article{fowler2012surface,
  title   = {Surface codes: Towards practical large-scale quantum computation},
  author  = {Fowler, Austin G. and Mariantoni, Matteo and Martinis, John M. and Cleland, Andrew N.},
  journal = {Physical Review A},
  volume  = {86},
  number  = {3},
  pages   = {032324},
  year    = {2012},
  publisher = {American Physical Society}
}

@article{horsman2012surface,
  title   = {Surface code quantum computing by lattice surgery},
  author  = {Horsman, Dominic and Fowler, Austin G. and Devitt, Simon and Van Meter, Rodney},
  journal = {New Journal of Physics},
  volume  = {14},
  number  = {12},
  pages   = {123011},
  year    = {2012},
  publisher = {IOP Publishing}
}

@article{Weiden_2025,
   title={High-Precision Multi-Qubit Clifford+T Synthesis by Unitary Diagonalization},
   volume={426},
   ISSN={2075-2180},
   url={http://dx.doi.org/10.4204/EPTCS.426.8},
   DOI={10.4204/eptcs.426.8},
   journal={Electronic Proceedings in Theoretical Computer Science},
   publisher={Open Publishing Association},
   author={Weiden, Mathias and Kalloor, Justin and Kubiatowicz, John and Younis, Ed and Iancu, Costin},
   year={2025},
   month=Aug, pages={215–230} }

@article{
    paradis_2024_synthetiq,
    author = {Paradis, Anouk and Dekoninck, Jasper and Bichsel, Benjamin and Vechev, Martin},
    title = {Synthetiq: Fast and Versatile Quantum Circuit Synthesis},
    year = {2024},
    issue_date = {April 2024},
    publisher = {Association for Computing Machinery},
    address = {New York, NY, USA},
    volume = {8},
    number = {OOPSLA1},
    url = {https://doi.org/10.1145/3649813},
    doi = {10.1145/3649813},
    journal = {Proc. ACM Program. Lang.},
    month = apr,
    articleno = {96},
    numpages = {28}
}

@inproceedings{silva2024lssp,
  author    = {Silva, Allyson and Zhang, Xiangyi and Webb, Zak and Kramer, Mia and Yang, Chan-Woo and Liu, Xiao and Lemieux, Jessica and Chen, Ka-Wai and Scherer, Artur and Ronagh, Pooya},
  title     = {Multi-qubit Lattice Surgery Scheduling},
  booktitle = {19th Conference on the Theory of Quantum Computation, Communication and Cryptography (TQC 2024)},
  series    = {Leibniz International Proceedings in Informatics (LIPIcs)},
  volume    = {310},
  pages     = {1:1--1:24},
  publisher = {Schloss Dagstuhl -- Leibniz-Zentrum f{\"u}r Informatik},
  year      = {2024},
  doi       = {10.4230/LIPIcs.TQC.2024.1},
  url       = {https://drops.dagstuhl.de/entities/document/10.4230/LIPIcs.TQC.2024.1},
}

@article{Litinski_2019,
   title={A Game of Surface Codes: Large-Scale Quantum Computing with Lattice Surgery},
   volume={3},
   ISSN={2521-327X},
   url={http://dx.doi.org/10.22331/q-2019-03-05-128},
   DOI={10.22331/q-2019-03-05-128},
   journal={Quantum},
   publisher={Verein zur Forderung des Open Access Publizierens in den Quantenwissenschaften},
   author={Litinski, Daniel},
   year={2019},
   month=mar, pages={128}
}

@article{Campbell_Synthillation,
  title = {Unified framework for magic state distillation and multiqubit gate synthesis with reduced resource cost},
  author = {Campbell, Earl T. and Howard, Mark},
  journal = {Phys. Rev. A},
  volume = {95},
  issue = {2},
  pages = {022316},
  numpages = {22},
  year = {2017},
  month = {Feb},
  publisher = {American Physical Society},
  doi = {10.1103/PhysRevA.95.022316},
  url = {https://link.aps.org/doi/10.1103/PhysRevA.95.022316}
}

@misc{gidney2024cultivation,
      title={Magic state cultivation: growing T states as cheap as CNOT gates},
      author={Craig Gidney and Noah Shutty and Cody Jones},
      year={2024},
      eprint={2409.17595},
      archivePrefix={arXiv},
      primaryClass={quant-ph},
      url={https://arxiv.org/abs/2409.17595},
}

@article{transgate,
  title = {Restrictions on Transversal Encoded Quantum Gate Sets},
  author = {Eastin, Bryan and Knill, Emanuel},
  journal = {Phys. Rev. Lett.},
  volume = {102},
  issue = {11},
  pages = {110502},
  numpages = {4},
  year = {2009},
  month = {Mar},
  publisher = {American Physical Society},
  doi = {10.1103/PhysRevLett.102.110502},
  url = {https://link.aps.org/doi/10.1103/PhysRevLett.102.110502}
}

@article{Gottesman_1999,
   title={Demonstrating the viability of universal quantum computation using teleportation and single-qubit operations},
   volume={402},
   ISSN={1476-4687},
   url={http://dx.doi.org/10.1038/46503},
   DOI={10.1038/46503},
   number={6760},
   journal={Nature},
   publisher={Springer Science and Business Media LLC},
   author={Gottesman, Daniel and Chuang, Isaac L.},
   year={1999},
   month=Nov, pages={390–393} }

@book{kitaev2002classical,
  author    = {Kitaev, Alexei Yu. and Shen, Alexander H. and Vyalyi, Mikhail N.},
  title     = {Classical and Quantum Computation},
  publisher = {American Mathematical Society},
  year      = {2002},
  edition   = {1},
}

@misc{
    landahl_2013_cisc,
	title = {Complex instruction set computing architecture for performing accurate quantum \${Z}\$ rotations with less magic},
	url = {http://arxiv.org/abs/1302.3240},
	doi = {10.48550/arXiv.1302.3240},
	urldate = {2024-07-03},
	author = {Landahl, Andrew J. and Cesare, Chris},
	month = oct,
	year = {2013},
	note = {arXiv:1302.3240 [quant-ph]},
}

@article{
    bocharov_2015_rus,
	title = {Efficient synthesis of universal {Repeat}-{Until}-{Success} circuits},
	volume = {114},
	issn = {0031-9007, 1079-7114},
	url = {http://arxiv.org/abs/1404.5320},
	doi = {10.1103/PhysRevLett.114.080502},
	number = {8},
	urldate = {2024-07-28},
	journal = {Physical Review Letters},
	author = {Bocharov, Alex and Roetteler, Martin and Svore, Krysta M.},
	month = feb,
	year = {2015},
	note = {arXiv:1404.5320 [quant-ph]},
	pages = {080502},
}

@article{
    bocharov_2015_fallback,
	title = {Efficient synthesis of probabilistic quantum circuits with fallback},
	volume = {91},
	issn = {1050-2947, 1094-1622},
	url = {http://arxiv.org/abs/1409.3552},
	doi = {10.1103/PhysRevA.91.052317},
	number = {5},
	urldate = {2024-07-28},
	journal = {Physical Review A},
	author = {Bocharov, Alex and Roetteler, Martin and Svore, Krysta M.},
	month = may,
	year = {2015},
	note = {arXiv:1409.3552 [quant-ph]},
	pages = {052317},
}

@INPROCEEDINGS{Chen_holistic,
  author={Chen, Tian-Fu and Liu, Cheng-Han and Jiang, Jie-Hong R.},
  booktitle={2024 IEEE International Conference on Quantum Computing and Engineering (QCE)}, 
  title={A Holistic Approach to Rotation Synthesis for Fault- Tolerant Quantum Computation}, 
  year={2024},
  volume={01},
  number={},
  pages={1026-1036},
  doi={10.1109/QCE60285.2024.00122}}

@article{Duclos_Cianci_2015,
   title={Reducing the quantum-computing overhead with complex gate distillation},
   volume={91},
   ISSN={1094-1622},
   url={http://dx.doi.org/10.1103/PhysRevA.91.042315},
   DOI={10.1103/physreva.91.042315},
   number={4},
   journal={Physical Review A},
   publisher={American Physical Society (APS)},
   author={Duclos-Cianci, Guillaume and Poulin, David},
   year={2015},
   month=Apr }

@misc{vaknin2025efficientmagicstatecultivation,
      title={Efficient Magic State Cultivation on the Surface Code},
      author={Yotam Vaknin and Shoham Jacoby and Arne Grimsmo and Alex Retzker},
      year={2025},
      eprint={2502.01743},
      archivePrefix={arXiv},
      primaryClass={quant-ph},
      url={https://arxiv.org/abs/2502.01743},
}

@misc{sahay2025foldtransversalsurfacecodecultivation,
      title={Fold-transversal surface code cultivation},
      author={Kaavya Sahay and Pei-Kai Tsai and Kathleen Chang and Qile Su and Thomas B. Smith and Shraddha Singh and Shruti Puri},
      year={2025},
      eprint={2509.05212},
      archivePrefix={arXiv},
      primaryClass={quant-ph},
      url={https://arxiv.org/abs/2509.05212},
}

@article{steane1996qec,
  author = {Steane, Andrew},
  year = {1996},
  month = {01},
  pages = {},
  title = {Multiple Particle Interference and Quantum Error Correction},
  volume = {452},
  journal = {Proceedings of the Royal Society A: Mathematical, Physical and Engineering Sciences},
  doi = {10.1098/rspa.1996.0136}
}

@misc{hofmeyr2026,
      title={Scheduling Lattice Surgery with Magic State Cultivation}, 
      author={Steven Hofmeyr and Mathias Weiden and Justin Kalloor and John Kubiatowicz and Costin Iancu},
      year={2026},
      eprint={2512.06484},
      archivePrefix={arXiv},
      primaryClass={quant-ph},
      url={https://arxiv.org/abs/2512.06484}, 
}

@article{CodyJones_2012,
doi = {10.1088/1367-2630/14/11/115023},
url = {https://doi.org/10.1088/1367-2630/14/11/115023},
year = {2012},
month = {nov},
publisher = {IOP Publishing},
volume = {14},
number = {11},
pages = {115023},
author = {Cody Jones, N and Whitfield, James D and McMahon, Peter L and Yung, Man-Hong and Meter, Rodney Van and Aspuru-Guzik, Alán and Yamamoto, Yoshihisa},
title = {Faster quantum chemistry simulation on fault-tolerant quantum computers},
journal = {New Journal of Physics}
}

@misc{bqskit,
title = {Berkeley Quantum Synthesis Toolkit (BQSKit) v1},
author = {Younis, Ed and Iancu, Costin C. and Lavrijsen, Wim and Davis, Marc and Smith, Ethan},
abstractNote = {The Berkeley Quantum Synthesis Toolkit (BQSKit) is an optimizing quantum compiler and research vehicle that combines ideas from several projects at LBNL into one easily accessible and quickly extensible software package. The ideas in the QFAST, QSearch, LEAP, and QFactor software tools (all licensed through ipo.lbl.gov) all build upon one another. By combining these into one package, we create symbiotic interactions between the tools. This means better results, better throughput, less to maintain, and greater surface area to the public. Additionally, the BQSKit tool will create a research platform for future work here at LBNL.},
doi = {10.11578/dc.20210603.2},
url = {https://doi.org/10.11578/dc.20210603.2},
howpublished = {[Computer Software] \url{https://doi.org/10.11578/dc.20210603.2}},
year = {2021},
month = {apr}
}

@phdthesis{ntro,
author = {Davis, Marc Grau},
school={Massachusetts Institute of Technology},
year = {2023},
month = {02},
pages = {},
title = {Numerical Synthesis of Arbitrary Multi-Qubit Unitaries with low T-Count},
doi = {10.13140/RG.2.2.32303.30886}
}

@misc{qiskit,
      title={Quantum computing with {Q}iskit},
      author={Javadi-Abhari, Ali and Treinish, Matthew and Krsulich, Kevin and Wood, Christopher J. and Lishman, Jake and Gacon, Julien and Martiel, Simon and Nation, Paul D. and Bishop, Lev S. and Cross, Andrew W. and Johnson, Blake R. and Gambetta, Jay M.},
      year={2024},
      doi={10.48550/arXiv.2405.08810},
      eprint={2405.08810},
      archivePrefix={arXiv},
      primaryClass={quant-ph}
}

@article{tket,
	title = {tket : {A} {Retargetable} {Compiler} for {NISQ} {Devices}},
	volume = {6},
	issn = {2058-9565},
	shorttitle = {t\${\textbar}\$ket\${\textbackslash}rangle\$},
	url = {http://arxiv.org/abs/2003.10611},
	doi = {10.1088/2058-9565/ab8e92},
	number = {1},
	urldate = {2023-11-13},
	journal = {Quantum Science and Technology},
	author = {Sivarajah, Seyon and Dilkes, Silas and Cowtan, Alexander and Simmons, Will and Edgington, Alec and Duncan, Ross},
	month = jan,
	year = {2021},
	note = {arXiv:2003.10611 [quant-ph]},
	pages = {014003},
}

@misc{ross_sellinger,
      title={Optimal ancilla-free Clifford+T approximation of z-rotations}, 
      author={Neil J. Ross and Peter Selinger},
      year={2016},
      eprint={1403.2975},
      archivePrefix={arXiv},
      primaryClass={quant-ph},
      url={https://arxiv.org/abs/1403.2975}, 
}

@misc{farhi_quantum_2014,
	title = {A {Quantum} {Approximate} {Optimization} {Algorithm}},
	url = {http://arxiv.org/abs/1411.4028},
	doi = {10.48550/arXiv.1411.4028},
	urldate = {2023-11-13},
	publisher = {arXiv},
	author = {Farhi, Edward and Goldstone, Jeffrey and Gutmann, Sam},
	month = nov,
	year = {2014},
	note = {arXiv:1411.4028 [quant-ph]},
}

@article{Gidney2018halvingcostof,
  doi = {10.22331/q-2018-06-18-74},
  url = {https://doi.org/10.22331/q-2018-06-18-74},
  title = {Halving the cost of quantum addition},
  author = {Gidney, Craig},
  journal = {{Quantum}},
  issn = {2521-327X},
  publisher = {{Verein zur F{\"{o}}rderung des Open Access Publizierens in den Quantenwissenschaften}},
  volume = {2},
  pages = {74},
  month = jun,
  year = {2018}
}

@article{Sanders_Combinatorial,
  title = {Compilation of Fault-Tolerant Quantum Heuristics for Combinatorial Optimization},
  author = {Sanders, Yuval R. and Berry, Dominic W. and Costa, Pedro C.S. and Tessler, Louis W. and Wiebe, Nathan and Gidney, Craig and Neven, Hartmut and Babbush, Ryan},
  journal = {PRX Quantum},
  volume = {1},
  issue = {2},
  pages = {020312},
  numpages = {70},
  year = {2020},
  month = {Nov},
  publisher = {American Physical Society},
  doi = {10.1103/PRXQuantum.1.020312},
  url = {https://link.aps.org/doi/10.1103/PRXQuantum.1.020312}
}

@article{cerezo_challenges_2022,
	title = {Challenges and opportunities in quantum machine learning},
	volume = {2},
	copyright = {2022 Springer Nature America, Inc.},
	issn = {2662-8457},
	url = {https://www.nature.com/articles/s43588-022-00311-3},
	doi = {10.1038/s43588-022-00311-3},
	language = {en},
	number = {9},
	urldate = {2023-11-13},
	journal = {Nature Computational Science},
	author = {Cerezo, M. and Verdon, Guillaume and Huang, Hsin-Yuan and Cincio, Lukasz and Coles, Patrick J.},
	month = sep,
	year = {2022},
	note = {Number: 9
Publisher: Nature Publishing Group},
	pages = {567--576},
}

@article{rahman_2022,
	title = {Self-mitigating {Trotter} circuits for {SU}(2) lattice gauge theory on a quantum computer},
	volume = {106},
	issn = {2470-0010, 2470-0029},
	url = {http://arxiv.org/abs/2205.09247},
	doi = {10.1103/PhysRevD.106.074502},
	number = {7},
	urldate = {2025-10-02},
	journal = {Physical Review D},
	author = {Rahman, Sarmed A. and Lewis, Randy and Mendicelli, Emanuele and Powell, Sarah},
	month = oct,
	year = {2022},
	note = {arXiv:2205.09247 [hep-lat]},
	pages = {074502},
}

@article{gilchrist_distance_2005,
    title = {Distance measures to compare real and ideal quantum processes},
    volume = {71},
    issn = {1050-2947, 1094-1622},
    url = {http://arxiv.org/abs/quant-ph/0408063},
    doi = {10.1103/PhysRevA.71.062310},
    number = {6},
    urldate = {2022-09-29},
    journal = {Physical Review A},
    author = {Gilchrist, Alexei and Langford, Nathan K. and Nielsen, Michael A.},
    month = jun,
    year = {2005},
    note = {arXiv:quant-ph/0408063},
    pages = {062310},
}

\end{document}